\documentclass[twocolumn]{article}

\usepackage[dvips]{graphicx}
\usepackage{amssymb}
\usepackage{float}

\def\eps{\varepsilon}

\title{Breakdown of adiabatic invariance in spherical tokamaks}

\author{Johan Carlsson\\[3mm]
Oak Ridge National Laboratory, P.O.~Box 2009, Oak Ridge, TN 37831--8071, USA}

\begin{document}

\maketitle

{\small ABSTRACT\@.
Thermal ions in spherical tokamaks have two adiabatic invariants:
the magnetic moment and the longitudinal invariant. For hot ions,
variations in magnetic-field strength over a gyro period can become
sufficiently large to cause breakdown of the adiabatic invariance.
The magnetic moment is more sensitive to perturbations than the 
longitudinal invariant and  there exists an intermediate regime, 
super-adiabaticity, where the longitudinal invariant remains adiabatic,
but the magnetic moment does not. The motion of super-adiabatic ions
remains integrable and confinement is thus preserved. However, above a
threshold energy, the longitudinal invariant becomes non-adiabatic too,
and confinement is lost as the motion becomes chaotic. We predict beam ions
in  present-day spherical tokamaks to be super-adiabatic but fusion alphas
in proposed burning-plasma spherical tokamaks to be non-adiabatic.}\\

Consider a charged particle, with mass $m$ and charge $q$, gyrating with speed
$v$ in a magnetic field, with field strength $B$. Assume axi-symmetric toroidal
geometry with coordinates ($R$, $\phi$, $Z$). The motion preserves the energy
$E$ (because the Lorentz force performs no work) and canonical toroidal
momentum $P_\phi$ (because the toroidal coordinate is ignorable).
There is no third exact constant of motion, but the normalized magnetic moment
\begin{displaymath}
\Lambda = \frac{B_0 \mu}{E} = \frac{B_0 v_{\perp}^2}{B \, v^2} \, ,
\end{displaymath}
where $v_{\perp}$ is the velocity component perpendicular to the magnetic
field ($v_{\perp}^2 = v^2 - v_{\parallel}^2$,
$v_{\parallel} = \mathbf{v} \cdot \mathbf{B} / B$)
and $B_0$ is the on-axis magnetic field, is an adiabatic invariant,
i.e.~as long as the variation in the magnetic field experienced by the
particle during one gyro period is small, $\Lambda$ oscillates at the gyro
frequency around a constant average. If we introduce the adiabaticity
parameter, $\eps$, the condition of slow (adiabatic) variation becomes
\begin{displaymath}
\eps = \frac{\varrho \, |\nabla{}B|}{B} \ll 1 \, ,
\end{displaymath}
where the gyro radius $\varrho = v_{\perp} / \Omega$, with the gyro frequency
$\Omega = |q| B / m$. The approximate constancy of the magnetic moment was
first pointed out by Alfv\'en~\cite{alfven}. Just as the magnetic moment is
associated with the gyro motion, a second adiabatic invariant is
associated with the slower drift motion: the longitudinal invariant
$J_\parallel = \oint{}\!v_\parallel d\ell$~\cite{longitudinal}.\\

When $\eps \ll 1$, there are thus three constants of motion, one for each
degree of freedom, i.e.~the motion is integrable, the orbits are quasi-periodic
and, in the absence of collisions, a particle will be eternally confined.
As the perturbation grows and $\eps \lesssim 1$, a dramatic change in the
character of the motion, from bounded to unbounded, will occur, as the
adiabaticity breaks down, the third constant of motion is lost, the motion
becomes non-integrable, and the orbits become chaotic. This process was first
investigated numerically~\cite{numerical} and later fully explained by
the rigorous KAM (Kolmogorov, Arnold \& Moser) theory~\cite{kam}.
Subsequently, Chirikov provided an approximate, but more practical, approach
when he introduced his semi-empirical resonance-overlap criterion and the
standard mapping~\cite{chirikov}. In the Chirikovian model, the breakdown
of adiabaticity is caused by strong nonlinear resonance between the
periodic motions occurring on different timescales.\newpage

\begin{figure}[H]
\begin{center}
\includegraphics[angle=90,scale=0.5]{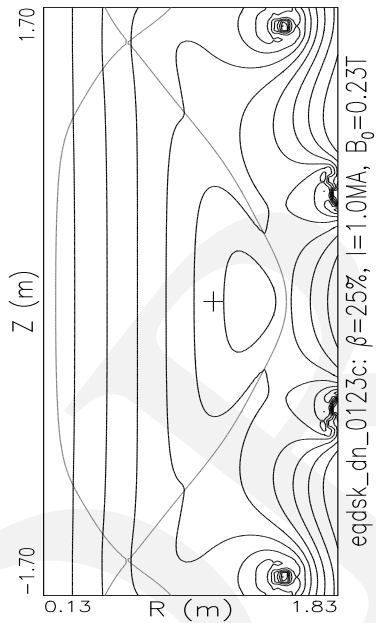}
\end{center}
\caption{25\% $\beta$ NSTX equilibrium.
Adjacent contours are equidistant in $1 / B$.
The thin, gray line is the separatrix and the cross marks the magnetic axis.
The poloidal-field coils, carrying approximately 200 kA of current each,
are easily identifiable to the right.}
\label{fig:contours}
\end{figure}

The compact and low-field spherical tokamak was introduced by Peng and
Strickler~\cite{st}. The first experiment, START~\cite{start}, showed
encouraging results and has been succeeded by two larger machines,
MAST~\cite{mast} and NSTX~\cite{nstx}. Larger, burning-plasma, experiments
have also been proposed~\cite{vns}.
The $B$ contours of spherical tokamaks are characterized by a magnetic
well, i.e.~a local minimum in $B$, centered outside of the magnetic axis
[see Fig.~\ref{fig:contours}]. For the case shown here, $B_{min} = 0.21$~T
and $B_{max} = 1.26$~T. The low value and large variation of $B$ in the plasma
(a factor of six) leads to a low threshold energy for the onset of
non-adiabatic motion. Furthermore, the threshold energy will vary substantially
over the cross section. We will demonstrate the consequences of this by
solving the Lorentz equation of motion for 80 kV deuterium ions moving
in the NSTX magnetic equilibrium shown in Fig.~\ref{fig:contours}.

\begin{figure}[!b]
\begin{center}
\includegraphics[angle=90,scale=0.5]{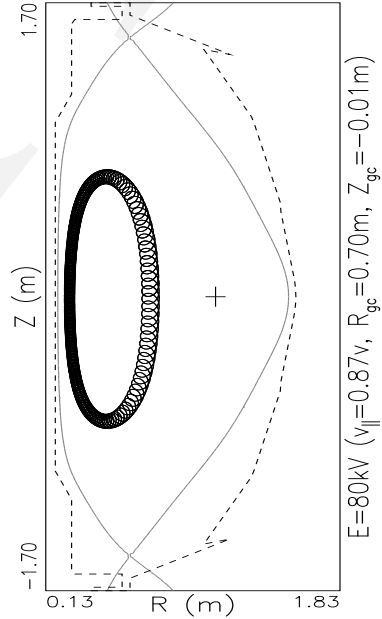}
\end{center}
\caption{Quasi-periodic orbit of a counter-injected deuterium beam ion.
The dashed line is the limiter.}
\label{fig:gyro_adiabatic}
\end{figure}

\begin{figure*}
\begin{center}
\includegraphics[angle=90,width=1.8\columnwidth]{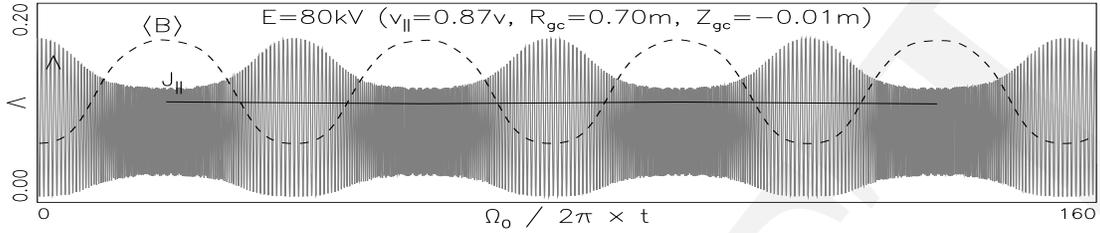}
\end{center}
\caption{Adiabatic time evolution of normalized magnetic moment $\Lambda$
(thin, gray line), longitudinal invariant $J_\parallel$ (solid line),
and gyro-averaged magnetic field strength $\langle B \rangle$ (dashed line)
for the orbit shown in Fig.~\ref{fig:gyro_adiabatic}.}
\label{fig:mu_adiabatic}
\end{figure*}

At NSTX, co-injection of an 80~kV deuterium beam is used to heat the plasma
(co-injection is favored over counter-injection because it results in lower
prompt losses)~\cite{nstx}. Note that the beam directions are relative to the
plasma current, which is in the opposite direction of the toroidal magnetic
field, and a co-injected ion thus has $v_{\parallel} < 0$. Here, however,
we will first launch a counter-injected ion, born with its gyro center just
below the midplane at $R_{gc}$ = 0.70~m. The resulting orbit is shown in
Fig.~\ref{fig:gyro_adiabatic}.
Plotting the time evolution of the normalized magnetic moment $\Lambda$
[see Fig.~\ref{fig:mu_adiabatic}], the adiabaticity is evident; $\Lambda$
oscillates around the constant mean with the amplitude spiking when the
ion passes through the minimum of $B$ along the orbit.
The longitudinal invariant $J_\parallel$ is practically constant.\\

\begin{figure}[H]
\begin{center}
\includegraphics[angle=90,scale=0.5]{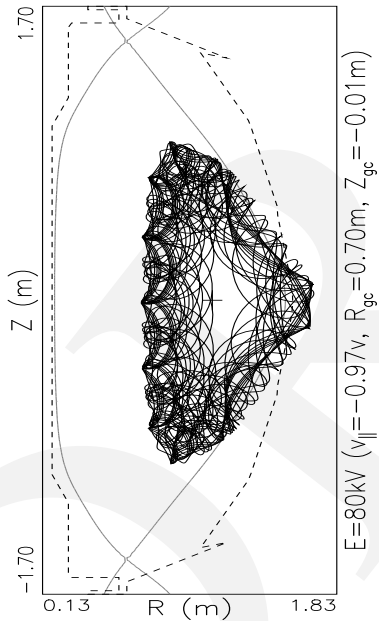}
\end{center}
\caption{Chaotic orbit of a co-injected beam ion.}
\label{fig:gyro_non-adiabatic}
\end{figure}

\begin{figure*}
\begin{center}
\includegraphics[angle=90,width=1.8\columnwidth]{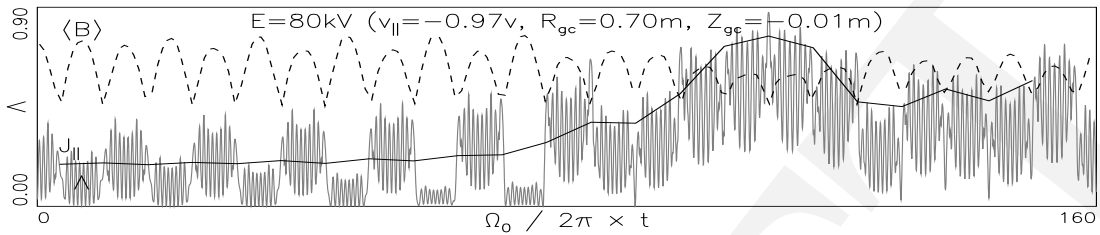}
\end{center}
\caption{Non-adiabatic time evolution of $\Lambda$ and $J_\parallel$ for
the orbit shown in Fig.~\ref{fig:gyro_non-adiabatic}.
$\Lambda$ exhibits characteristic, erratic jumps as $\langle B \rangle$
passes through local minima.}
\label{fig:mu_non-adiabatic}
\end{figure*}

Next, we launch a co-injected ion with its gyro center at exactly the same
location as before for the counter injection. As can be seen in
Fig.~\ref{fig:gyro_non-adiabatic}, the solution to the equation of motion
now takes on a completely different character, as the adiabaticity condition
is violated.
The variation of the magnetic field along the orbit is actually less than
it was for the counter-injected ion [see Fig.~\ref{fig:mu_non-adiabatic}];
but because the gyro period is longer and the drift period is shorter,
the change over one gyration becomes sufficiently large to destroy the
\enlargethispage*{-1ex}
adiabaticity of both $\Lambda$ and $J_\parallel$.\\

\begin{figure}[H]
\begin{center}
\includegraphics[angle=90,scale=0.5]{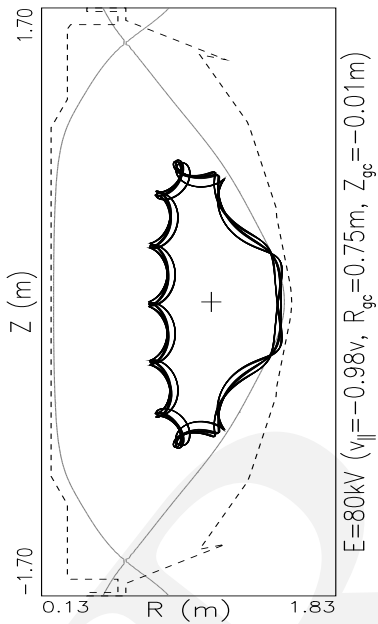}
\end{center}
\caption{Almost quasi-periodic orbit of a super-adiabatic co-injected beam
ion on the edge of the magnetic well.}
\label{fig:gyro_super-adiabatic}
\end{figure}

\begin{figure*}
\begin{center}
\includegraphics[angle=90,width=1.8\columnwidth]{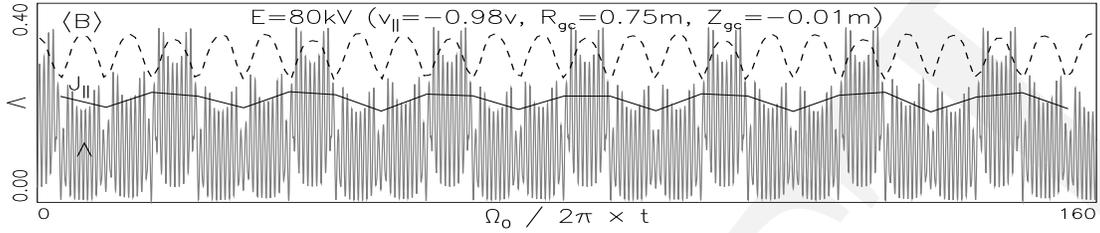}
\end{center}
\caption{Non-adiabatic time evolution of $\Lambda$ but adiabatic time
evolution of $J_\parallel$ for the orbit shown in
Fig.~\ref{fig:gyro_super-adiabatic}.}
\label{fig:mu_super-adiabatic}
\end{figure*}

We now move the gyro center five centimeters outward, to the edge of the
magnetic well.
In the magnetic well the $B$ contours are nested, closed loops and roughly
overlap the flux surfaces. The variation of $B$ along a flux surface is thus
greatly reduced. Moving closer to the magnetic well, the adiabaticity
parameter will gradually decrease, and launching the ion at $R_{gc}$ = 0.75~m,
the character of the motion is again changed
[see Fig.~\ref{fig:gyro_super-adiabatic}].
Notice how the ion shifts between three different quasi-periodic orbits.
This behavior can be understood by inspecting the time evolution of
$\Lambda$ and $J_\parallel$ [see Fig.~\ref{fig:mu_super-adiabatic}].
As can be seen, $\Lambda$ does not oscillate around a constant average
and is thus non-adiabatic. Instead it jumps between three different gyro
averages, explaining the peculiar ion orbit. In contrast, $J_\parallel$
oscillates with a period of three drift periods around a constant
average and thus qualifies as an adiabatic invariant. This phenomenon,
with non-adiabatic $\Lambda$ but adiabatic $J_\parallel$ is known as
super-adiabaticity~\cite{super}. Note that in the
super-adiabatic regime, the longitudinal invariant $J_\parallel$ replaces
the magnetic moment as the third constant of motion, and the motion remains
integrable. The confinement of super-adiabatic ions should thus be similar
to that of adiabatic ions. As the adiabaticity parameter is further reduced
and we leave the super-adiabatic regime, however, only two
constants of motion will remain, and the confinement will be drastically
decreased as the motion becomes chaotic.\\

The non-adiabatic beam ion shown in Fig.~\ref{fig:gyro_non-adiabatic}
intersects the limiter, and it would have done so even if it had been
adiabatic. This seems to be typical for NSTX: the non-adiabaticity only
affects 80 kV beam ions already following loss orbits and should thus not
have any substantial, negative effect on the confinement. The majority of
the confined beam ions are super-adiabatic, similar to the orbit in
Fig.~\ref{fig:gyro_super-adiabatic}.
\enlargethispage*{6ex} In fact, even orbits deep inside the well
have a slightly non-adiabatic magnetic moment.\\

In START, co-injection of a 30 kV hydrogen beam was found to have
$\sim$60\% absorption efficiency~\cite{akers}. Based on numerical simulations,
these beam ions were claimed to be strongly non-adiabatic. This would seem
inconsistent with the good confinement and we note that Fig.~3a of
Ref.~\cite{akers}, which shows the time evolution of the magnetic moment for
a typical beam ion and is said to demonstrate the non-adiabaticity, in fact
indicates super-adiabatic, or possibly even fully adiabatic, beam ions.\\

\begin{figure*}
\begin{center}
\includegraphics[angle=90,width=1.8\columnwidth]{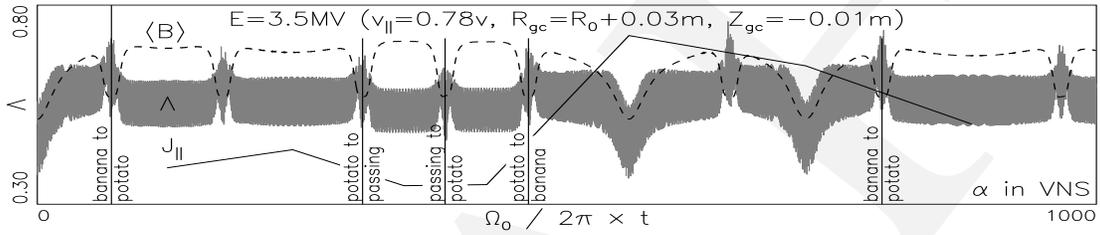}
\end{center}
\caption{Non-adiabatic time evolution of $\Lambda$ and $J_\parallel$ for
fusion $\alpha$ in VNS.}
\label{fig:mu_VNS}
\end{figure*}

If the beam energy is increased to 110 kV in NSTX, the situation remains
essentially the same: the vast majority of the beam ions that are not
promptly lost, are super-adiabatic, and their motion is consequently bounded.
However, in future, burning-plasma experiments, such as the VNS~\cite{vns},
the fusion-born $\alpha$-particles will display the whole range of dynamics:
adiabatic, super-adiabatic, and non-adiabatic. As an example, we show the
time evolution of the magnetic moment for an alpha, born just below the
midplane and three centimeters outside of the magnetic axis, for a
38\% $\beta$ VNS equilibrium [see Fig.~\ref{fig:mu_VNS}]. In addition to
possible direct non-adiabatic losses when a giant jump in $\Lambda$ causes
transition to a loss orbit, there is also an indirect loss mechanism.
As can be seen, the jumps in $\Lambda$ force the $\alpha$ back and forth
across the trapped-passing boundary. When driven by ion-ion collisions,
this is of course the process that generates neo-classical transport.
For fusion-born $\alpha$-particles, the drift period is many orders of
\enlargethispage*{3ex}
magnitude shorter than the collisional timescale, and the
non-adiabatic transport mechanism described above can thus be expected to
affect the overall $\alpha$ confinement. Yet another effect to consider is
the Arnold diffusion~\cite{arnold}, which might be important if the
magnetic-field ripple is large enough to destroy the axi-symmetry.\\

\textbf{Acknowledgment:} This Letter has benefited greatly from the
insightful comments of Fred Jaeger. The magnetic-equilibrium data was
kindly provided by Dennis Strickler.
This research was sponsored by the Oak Ridge National Laboratory,
managed by UT--Battelle, LLC, for the U.S.~Department of Energy under
contract DE--AC05--00OR22725. Research was supported in part by an
appointment to  the ORNL Postdoctoral Research Associates Program,
administered jointly by Oak Ridge National Laboratory and the Oak Ridge
Institute for Science and Education.\\



\clearpage

\end{document}